\documentclass[pra, aps,  twocolumn,
showpacs]{revtex4-1}

\usepackage[T1]{fontenc}
\usepackage[colorlinks,urlcolor=blue,citecolor=blue,linkcolor=blue]{hyperref}

\usepackage{physics}
\usepackage[english]{babel} 
\usepackage[english]{layout}
\usepackage{amsmath,amsfonts,amssymb}
\usepackage{graphicx}
\usepackage{float}
\usepackage{color}
\usepackage[dvipsnames]{xcolor}
\usepackage{braket}
\usepackage{version}
\usepackage{array,multirow,makecell}
\usepackage{afterpage}
\usepackage[toc,page]{appendix} 
\usepackage{bbold}

\let\oldk\k
\renewcommand{\k}{{\bf k}}

\newcommand{\ob}[1]{{\color{black}#1}}

\newcommand{\moved}[1]{{\color{black}#1}}
\begin{document}

\title{Dissipative light-matter coupling and anomalous dispersion in nonideal cavities}

\author{Olivier Bleu}
\affiliation{School of Physics and Astronomy, Monash University, Victoria 3800, Australia}
\affiliation{ARC Centre of Excellence in Future Low-Energy Electronics Technologies, Monash University, Victoria 3800, Australia}

\author{Kenneth Choo}
\affiliation{School of Physics and Astronomy, Monash University, Victoria 3800, Australia}
\affiliation{ARC Centre of Excellence in Future Low-Energy Electronics Technologies, Monash University, Victoria 3800, Australia}

\author{Jesper Levinsen}
\affiliation{School of Physics and Astronomy, Monash University, Victoria 3800, Australia}
\affiliation{ARC Centre of Excellence in Future Low-Energy Electronics Technologies, Monash University, Victoria 3800, Australia}

\author{Meera M. Parish}
\affiliation{School of Physics and Astronomy, Monash University, Victoria 3800, Australia}
\affiliation{ARC Centre of Excellence in Future Low-Energy Electronics Technologies, Monash University, Victoria 3800, Australia}

\begin{abstract}

We consider the scenario of an emitter embedded in a nonideal cavity. 
Using an input-output approach to describe the open system, we show that an effective dissipative coupling  between the emitter and the cavity can emerge because of their interaction with a common photonic environment.
The underlying mechanism is independent of the nature of the emitter and exists even at zero temperature; hence our results provide a pathway for accessing a range of non-Hermitian phenomena in
a variety of light-matter coupled systems. 
In particular, we show that the effective dissipative coupling can lead to the phenomenon of level attraction between the emitter and cavity mode when
the radiative decay rates exceed the conventional Rabi coupling. 
Our model thus provides a possible explanation for the anomalous dispersions and negative mass 
observed in recent photoluminescence measurements in semiconductor microcavities.
Finally, we show that our effective non-Hermitian system can produce hybrid light-matter exceptional points and bound states in the continuum.

\end{abstract}

\maketitle

\section{Introduction}

In quantum optics, the spontaneous emission of radiation is understood as arising from the coupling between an emitter and the vacuum of the electromagnetic field in its surroundings~\footnote{Interestingly, the concept of spontaneous emission predates the formal quantization of the electromagnetic field as well as the introduction of the word \textit{photon}}. Importantly, this means that the spontaneous emission is not an intrinsic property of the emitter, but can be controlled (enhanced or inhibited) by placing the emitter in an optical cavity, thus modifying its electromagnetic environment \cite{purcell_spontaneous_1946,Goy1983,Kleppner1981,Yablonovitch1987}. This fact is at the heart of 
the research field of cavity quantum electrodynamics \cite{Walther_2006,Reiserer2015,RMPLodahl}. 

In the ideal case, there are negligible interactions between the cavity and the outside electromagnetic environment, allowing one  to achieve a strong coupling between an emitter and a cavity photon mode. Here, the eigenstates at low excitation density consist of superpositions between the emitter and bare cavity photon states --- the so-called polariton states. In practice, however, the cavity cannot be perfect and the light-matter-coupled system is affected by the external environment.  
In this context, there are two effects: 
the damping of the cavity, which has been extensively studied theoretically for the Jaynes-Cummings model~\cite{Sachdev1984,Barnett1986,Agarwal1986,Madajczyk1988}; 
and the possibility for the emitter itself to emit radiation outside of the cavity~\cite{Carmichael1989,Cirac1991,Auffeves2008,carmichael2009statisticalb}. 
Such dissipative effects are generic and have been similarly modelled for a range of scenarios beyond the original case of an atomic emitter, such as semiconductor microcavities with either two-level systems \cite{Andreani1999,Laucht2009,delValle2009,Giesz2016} or bosonic modes \cite{Verger2006,Laussy2008,Laussy2009} as emitters.
However, to our knowledge, these previous works have always assumed that the decay channels for the emitter and the cavity photon are independent.

\begin{figure}[tbp] 
    \includegraphics[width=\linewidth]{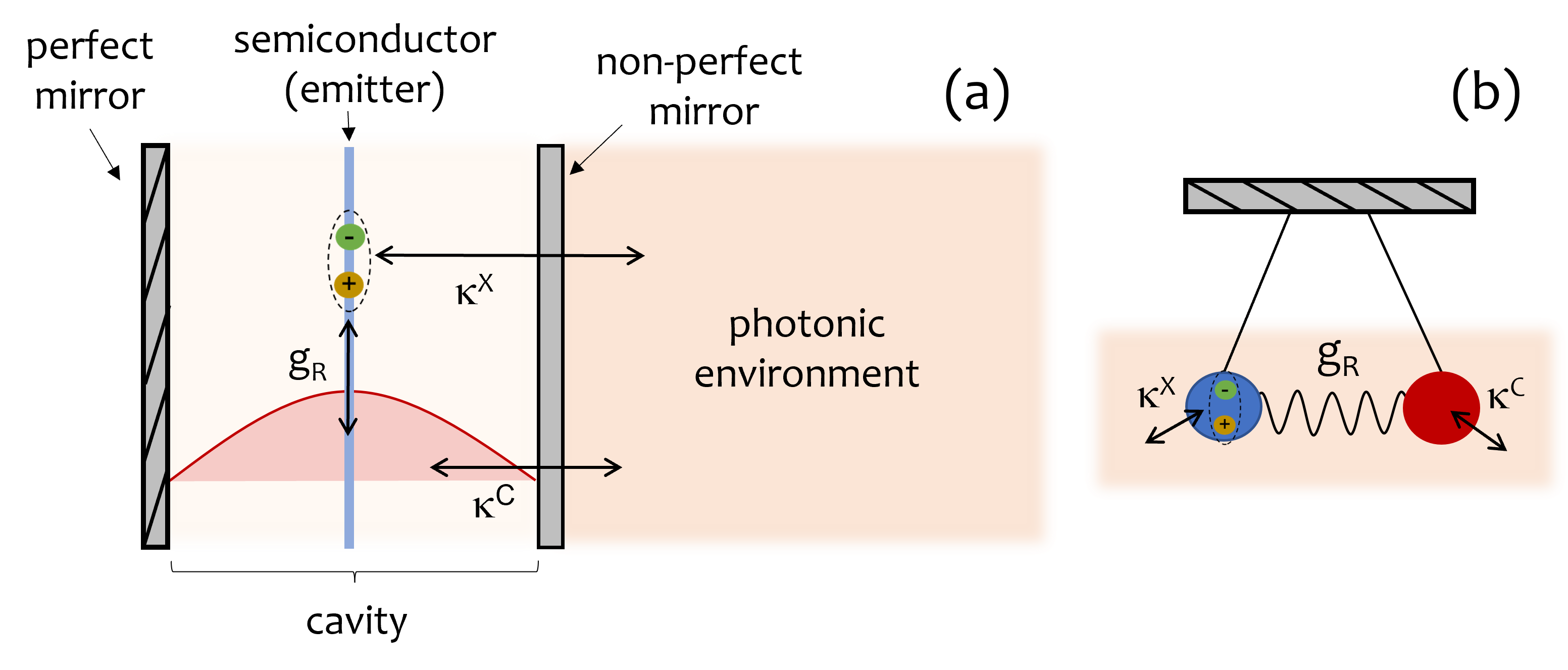}
\caption{(a) Sketch of the hybrid light-matter system under consideration. It consists of a semiconductor layer embedded in a planar optical cavity. One of the mirrors is not perfect, which allows the emitter and the cavity photon to interact with the common photonic environment via the couplings $\kappa$. Panel (b) illustrates the analogy with two coupled oscillators immersed in a common medium.}
\label{fig1}
\end{figure}

In this Article, we consider the situation in which the emitter and the cavity photon interact with a \textit{common} photonic environment, a scenario which is readily realized in a nonideal cavity [Fig.~\ref{fig1}(a)]. As we show, this gives rise to an effective dissipative coupling between the cavity photon and emitter, which is analogous to the induced interactions between two oscillators immersed in the same medium [Fig.~\ref{fig1}(b)]. 
In principle, this configuration only requires a mirror that is not perfect: it is independent of the nature of the emitter and exists even when the environment is at zero temperature. Therefore, 
\ob{our findings are of potential relevance for} 
a variety of experimental platforms.
To be concrete, we consider the system to be a planar semiconductor microcavity that hosts cavity photons and excitons \cite{kavokin2017microcavities}, and we use an input-output approach \cite{Gardiner1985,Ciuti2006} to describe the open quantum system. In particular, we find 
an effective dissipative light-matter coupling  
that can lead to level attraction between the exciton and cavity mode.
This effect 
provides a plausible and simple explanation for the recently observed anomalous dispersions  
in planar semiconductor microcavities 
\cite{Dhara2018,Wurdack2023}, as well as earlier reports of level attraction in quantum dot cavities \cite{Tawara2010,Dalacu2010,Valente2013}. 
We also demonstrate the existence of hybrid light-matter exceptional points (EPs) and bound states in the continuum (BiCs) in our model.

\ob{The paper is organized as follows. The model and formalism are introduced in Section~\ref{Sec:1}. In Section~\ref{Sec:2}, we present the results obtained by making use of the memoryless approximation. Finally, in Section~\ref{Sec:3} we extend the model to include non-radiative losses and calculate the reflection and absorption spectra. A brief summary and our conclusions are given in Section~\ref{Sec:Conc}. Additional information and technical details are provided in Appendices.}

\section{Theoretical description}\label{Sec:1}

\subsection{Model}
We employ a system-environment decomposition, as is customary for the description of open quantum systems, and
hence we start with a total Hamiltonian of the form $ \hat{H}=\hat{H}_S+\hat{H}_{E}+\hat{H}_{SE}$.
Here, 
$\hat{H}_S$ and $\hat{H}_{E}$ correspond to the 
system of interest and the environment, respectively, while $\hat{H}_{SE}$ 
describes the system-environment coupling. 
%
%
%
%
To be concrete, we consider the system illustrated in Fig.~\ref{fig1}(a).
It consists of two-dimensional excitons and cavity photons described by the Hamiltonian
\begin{equation}
    \hat{H}_S=\sum_\k\left[\epsilon_{\k}^C \hat{c}_{\k}^{\dagger} \hat{c}_{\k}+ \epsilon_{\k}^X \hat{x}_{\k}^{\dagger}\hat{x}_{\k}+g_R\left(\hat{x}_{\k}^{\dagger} \hat{c}_{\k}+\hat{c}_{\k}^{\dagger} \hat{x}_{\k}\right)\right].
\end{equation}
Here, $\hat{c}_{\mathbf{k}}$ ($\hat{c}_{\mathbf{k}}^\dagger$) and $\hat{x}_{\mathbf{k}}$ ($\hat{x}_{\mathbf{k}}^\dagger$) are bosonic annihilation (creation) operators of cavity photons and quantum-well excitons, respectively, with in-plane momentum $\hbar\mathbf{k}$. The kinetic energies at low momenta are $\epsilon_{\mathbf{k}}^C = \epsilon_0 +\hbar^2k^2/2m_C +\delta$ and  $\epsilon_{\mathbf{k}}^X = \epsilon_0+\hbar^2k^2/2m_X$, where $k \equiv |\k|$ and
$m_C$ ($m_X$) is the cavity photon (exciton) mass, while $\delta$ is the photon-exciton detuning and $\epsilon_0$ is the exciton transition energy.  
$g_R$ is the Rabi coupling, and we assume that the rotating wave approximation holds ($\epsilon_{0}\gg g_R$). 

We consider the environment to be the electromagnetic field outside of the cavity. To describe this photonic bath, we use the following Hamiltonian
\begin{equation}\label{eq:HamBathphot}
 \hat{H}_{E}= \int dq\sum_{\mathbf{k}} \epsilon_{\k,q}^{E} \hat{e}_{\k q}^\dagger \hat{e}_{\k q},
\end{equation}
where $\hat{e}_{\k q}$ ($ \hat{e}_{\k q}^\dagger$) are annihilation (creation) operators for photons outside the cavity with in- and out-of-plane wavevectors $\k$ and $q$, respectively, and $\epsilon_{\k,q}^{E}=\hbar c \sqrt{k^2+q^2}$ is the corresponding photon kinetic energy. 

Since we consider the situation where both the emitter and the cavity photons interact with the photonic environment, we describe the system-environment interaction with a Hamiltonian of the form $\hat{H}_{SE} =\hat{H}_{XE}+ \hat{H}_{CE}$ with
\begin{subequations}\label{eq:HamSE}
\begin{align}
  \hat{H}_{XE}&= \int dq\sum_{\mathbf{k}} \left( \kappa_{\k,q}^{X}  \hat{e}_{\k q}^\dagger \hat{x}_{\k}+ h.c.\right),\\
  \hat{H}_{CE}&= \int dq\sum_{\mathbf{k}} \left( \kappa_{\k,q}^{C}  \hat{e}_{\k q}^\dagger \hat{c}_{\k}+ h.c.\right) .
\end{align}
\end{subequations}
$\hat{H}_{XE}$ encodes the fact that the probability of an exciton to recombine by emitting a photon directly to the environment 
is nonzero \footnote{The nonzero coupling between the emitter and the photonic environment originates from the fact that when the mirror is not perfect the free space (untrapped) electromagnetic modes penetrate in the cavity.}.
In particular, this can become the dominant process for the radiative recombination of the emitter when the cavity mode and emitter frequencies are not resonant (i.e., the opposite of the Purcell effect~\cite{purcell_spontaneous_1946}). $\hat{H}_{CE}$ describes the cavity-environment coupling that exists even in the absence of the semiconductor layer.
Crucially, the form of $\hat{H}_{SE}$ allows for interference effects to take place when the couplings $\kappa^X_{\k,q}$ and $\kappa^C_{\k,q}$ are both nonzero. This is an essential difference between the present model and those which consider independent environments for the emitter and the cavity (e.g., Refs.~\cite{Carmichael1989,Cirac1991,Auffeves2008,carmichael2009statisticalb,Andreani1999,Laucht2009,delValle2009,Giesz2016,Verger2006,Laussy2008,Laussy2009}). 
For each in-plane wavevector, the scenario we consider is analogous to the problem of two coupled oscillators immersed in a common medium illustrated in Fig.~\ref{fig1}(b).

\subsection{Input-output formalism}

\subsubsection{Time dependent equations}
To obtain the time evolution of the system, we use an input-output approach \cite{Gardiner1985,Ciuti2006}, which allows us to describe the response to an arbitrary input such as a coherent field. The key idea is to start from the Heisenberg equations of motion for the system and bath operators, and then eliminate the bath operators by using their initial conditions, i.e., the input, \moved{as detailed below.

The Heisenberg equations of motion for the system operators are
\begin{subequations}\label{Eq.Heis_sys}
\begin{eqnarray} 
 i\hbar\partial_t\hat{c}_\k &=&\left[\hat{c}_\k,\hat{H}_S\right]+ \int dq\, \kappa_{\k,q}^{C*}\hat{e}_{\k q},\\ 
  i\hbar\partial_t\hat{x}_\k &=&\left[\hat{x}_\k,\hat{H}_S\right]+ \int dq\, \kappa_{\k,q}^{X*}\hat{e}_{\k q}.
\end{eqnarray}
\end{subequations}
In the same way, one obtains the evolution equations for the bath operators 
\begin{align} \label{Eq.Heis_bath}
 i\hbar\partial_t\hat{e}_{\k q} =&\epsilon_{\k,q}^{E} \hat{e}_{\k q} +  \kappa_{\k,q}^{C}\hat{c}_{\k}+\kappa_{\k,q}^{X}\hat{x}_{\k}.
\end{align}
The formal solution of Eq.~\eqref{Eq.Heis_bath} is of the form:
\begin{align} \label{Eq.bath_sol}
\hat{e}_{\k q}(t) &= e^{-\frac{i}{\hbar}\epsilon_{\k,q}^{E}(t-t_0)} \hat{e}_{\k q}(t_0)\\
& + \frac{1}{i\hbar}\int_{t_0}^t dt'  e^{-\frac{i}{\hbar}\epsilon_{\k,q}^{E}(t-t')}\left(\kappa_{\k,q}^{C}\hat{c}_{\k}(t')+\kappa_{\k,q}^{X} \hat{x}_{\k}(t')\right).\nonumber
\end{align}
If one defines input and output operators as
\begin{subequations}
\begin{align} 
\hat{e}_{\k q}^{\text{I}}&=\lim_{t_0\rightarrow-\infty}e^{\frac{i}{\hbar}\epsilon_{\k,q}^{E}t_0}\hat{e}_{\k q}(t_0) ,\\
\hat{e}_{\k q}^{\text{O}}&=\lim_{t\rightarrow\infty}e^{\frac{i}{\hbar}\epsilon_{\k,q}^{E}t}\hat{e}_{\k q}(t),
\end{align}
\end{subequations}
one can use Eq.~\eqref{Eq.bath_sol} to obtain the following relation
\begin{equation} \label{Eq.Inout1}
\hat{e}_{\k q}^{\text{O}}=\hat{e}_{\k q}^{\text{I}} +\frac{1}{i\hbar}\kappa_{\k,q}^{C}\hat{\mathcal{C}}_\k(\omega_{\k,q}^{E}) +\frac{1}{i\hbar}\kappa_{\k,q}^{X}\hat{\mathcal{X}}_\k(\omega_{\k,q}^{E})  .
\end{equation}
Here,
$\hat{\mathcal{A}}_\k(\omega)=\int dt\, e^{i\omega t}\hat{a}_\k(t)$ corresponds to the Fourier transform of the operator $\hat{a}_\k(t)$ and $\omega_{\k,q}^{E}=\epsilon_{\k,q}^{E}/\hbar$.
We can see that Eq.~\eqref{Eq.Inout1} relates the output operator to those of the input and the system.}

\moved{Inserting Eq.~\eqref{Eq.bath_sol} into Eq.~\eqref{Eq.Heis_sys} and taking the limit $t_0\rightarrow-\infty$, one obtains the following equations for the system operators}

\begin{widetext}
\begin{subequations}\label{Eq.Heis_Lang_main}
\begin{eqnarray} 
 i\hbar\partial_t\hat{c}_\k &=&\left[\hat{c}_\k,\hat{H}_S\right]- i \int_{-\infty}^{\infty} dt'\Gamma_\k^{CC}(t-t') \hat{c}_\k(t')  - i \int_{-\infty}^{\infty} dt'\Gamma_\k^{CX}(t-t') \hat{x}_\k(t')+\hat{F}_\k^{C}(t),\\ 
  i\hbar\partial_t\hat{x}_\k &=&\left[\hat{x}_\k,\hat{H}_S\right] - i \int_{-\infty}^{\infty} dt'\Gamma_\k^{XX}(t-t') \hat{x}_\k(t')  - i \int_{-\infty}^{\infty} dt'\Gamma_\k^{XC}(t-t') \hat{c}_\k(t')+\hat{F}_\k^{X}(t).
\end{eqnarray}
\end{subequations}
\end{widetext}
Here, both the Langevin-like force operators $\hat{F}_\k^{X,C}(t)$ and the damping terms $\Gamma_\k^{AB}(\tau)$ originate from the interaction with the common photonic environment \eqref{eq:HamSE} and are not captured in $\hat{H}_S$.
Importantly, we see that $\Gamma_\k^{CX}(\tau)$ in Eq.~\eqref{Eq.Heis_Lang_main} induces an additional coupling between the excitons and cavity photons.
\moved{Explicitly, one has
\begin{align}\label{Eq.gammaT}
\Gamma_\k^{AB}(\tau)&=\theta(\tau)\frac{1}{\hbar}\int dq\, \kappa_{\k,q}^{A*}\kappa_{\k,q}^{B}e^{-\frac{i}{\hbar}\epsilon_{\k,q}^{E}\tau},
\end{align}
and the `force' operators are related to the input operators as
\begin{subequations}
\begin{eqnarray} 
\hat{F}_\k^{C}(t)&=&\int dq\, \kappa_{\k,q}^{C*} e^{-\frac{i}{\hbar}\epsilon_{\k,q}^{E}t}\hat{e}_{\k q}^{\text{I}},
\\
\hat{F}_\k^{X}(t)&=&\int dq\, \kappa_{\k,q}^{X*} e^{-\frac{i}{\hbar}\epsilon_{\k,q}^{E}t}\hat{e}_{\k q}^{\text{I}}.
\end{eqnarray}
\end{subequations}
}%

\subsubsection{Frequency domain}
Taking the Fourier transform of Eq.~\eqref{Eq.Heis_Lang_main}, 
we arrive at the matrix equation
\begin{equation} \label{Eq.Heis_LangFour}
\mathcal{M}(\k,\omega)
\begin{pmatrix} 
\hat{\mathcal{C}}_\k(\omega)\\
\hat{\mathcal{X}}_\k(\omega)
 \end{pmatrix} = \begin{pmatrix} 
\hat{\mathcal{F}}_{\k}^C(\omega)\\
\hat{\mathcal{F}}_{\k}^X(\omega)
 \end{pmatrix} .
\end{equation}
Here, $\hat{\mathcal{X}}_\k(\omega),\hat{\mathcal{C}}_\k(\omega),\hat{\mathcal{F}}_\k(\omega)$ denote the Fourier transforms of the operators $\hat{x}_\k(t),\hat{c}_\k(t),\hat{F}_\k(t)$ respectively, and
\begin{align} \label{Eq.Mmat}
\setlength\arraycolsep{0pt}\mathcal{M}(\k,\omega)=\begin{pmatrix}
\hbar\omega-\epsilon_\k^C+i\tilde{\Gamma}_\k^{CC}(\omega)&&-g_R+i\tilde{\Gamma}_\k^{CX}(\omega)\\
-g_R+i\tilde{\Gamma}_\k^{XC}(\omega) && \hbar\omega-\epsilon_\k^X+i\tilde{\Gamma}_\k^{XX}(\omega)
\end{pmatrix}.
\end{align}
We note that ${\cal M}^{-1}$ %
corresponds to the single-particle Green's matrix of the system, modified by the interaction with the common photonic environment (see Appendix \ref{sec.AppBBA}).
 $\tilde{\Gamma}_\k^{AB}(\omega)$ denotes the Fourier transform of $\Gamma_\k^{AB}(\tau)$ \moved{and takes the form 
\begin{align} \label{Eq.Gamma_fourier}
\tilde{\Gamma}_\k^{AB}(\omega)&=\frac{\pi}{\hbar}\int dq \,  \kappa_{\k,q}^{A*}\kappa_{\k,q}^{B} \delta(\omega-\omega_{\k,q}^{E}) \nonumber \\
& ~~~~~~ +\frac{i}{\hbar}\mathcal{P}\int dq \frac{\kappa_{\k,q}^{A*}\kappa_{\k,q}^{B}}{\omega-\omega_{\k,q}^{E}},
\end{align}
where $\mathcal{P}$ denotes the Cauchy principal value. 
The Fourier transform of the force operators are related to the input operator as
\begin{subequations}\label{Eq.Force_fourier}
\begin{align} 
\hat{\mathcal{F}}_{\k}^{C}(\omega)&=2\pi\kappa_{\k,q_p}^{C*}\rho_{\k}(\omega)\hat{e}_{\k,q_p}^{\text{I}} , 
\\ 
\hat{\mathcal{F}}_{\k}^{X}(\omega)&=2\pi\kappa_{\k,q_p}^{X*}\rho_{\k}(\omega)\hat{e}_{\k,q_p}^{\text{I}},
\end{align}
\end{subequations}
where the wavevector $q_p$ is defined by the resonance condition $\omega=\omega_{\k,q_p}^{E}$ and we have introduced %
\begin{eqnarray} 
\rho_{\k}(\omega)=(\partial_q \omega_{\k,q}^{E})_{q=q_p}^{-1}.
\end{eqnarray}
Physically, $\rho_{\k}(\omega)$ corresponds to the photon environment density of states.
It is real and well defined when $\omega>c k$, while when $\omega<c k$, it is not possible to emit radiation outside of the cavity since there are no free photon modes.

Inserting Eqs.~\eqref{Eq.Heis_LangFour} and~\eqref{Eq.Force_fourier} in Eq.~\eqref{Eq.Inout1}, one obtains the following input-output relation
\begin{equation} \label{Eq.InoutF1}
\hat{e}_{\k,q_p}^{\text{O}}=
S(\k,\omega)
\hat{e}_{\k,q_p}^{\text{I}},
\end{equation}
where
\begin{align}
S(\k,\omega)=1-i\frac{2\pi\rho_{\k}(\omega)}{\hbar}\sum_{i,j}\left[\mathcal{M}(\k,\omega)\right]_{ij}^{-1} \kappa_{\k,q_p}^{i}\kappa_{\k,q_p}^{j*}. 
\end{align}
Here, $[\mathcal{M}(\k,\omega)]_{ij}^{-1}$ corresponds to the $(i,j)$th element of $\mathcal{M}^{-1}$.
One can check that $|S(\k,\omega)|=1$, and therefore the transformation \eqref{Eq.InoutF1} is unitary. This implies that the output operators obey the same Bose commutation relations as the input operators.  
However, it is interesting to remark that the fact that we have a single common photonic bath with a given density of state $\rho_{\k}$ is important for this unitarity property.
Indeed, in the different scenario where the emitter and cavity photon interact with independent matter and photon baths with different dispersion relations (i.e., different densities of states), the resulting input-output transformation cannot be unitary since it is not equivalent to have an input from the matter or from the photon bath. This subtlety seems to have been missed in Ref.~\cite{Ciuti2006}, and must be accounted for to calculate reflection or absorption, as discussed in Sec.~\ref{Sec:3}.}

While these observables cannot be properly defined in a model with a single environment, one can use Eq.~\eqref{Eq.Heis_LangFour} to calculate the power spectrum
 \begin{align} \label{Eq.Powerspect}
I(\k,\omega) =\langle \hat{\mathcal{C}}_\k^{\dagger}(\omega)\hat{\mathcal{C}}_\k(\omega)+ \hat{\mathcal{X}}_\k^{\dagger}(\omega)\hat{\mathcal{X}}_\k(\omega)\rangle ,
\end{align}
which encodes the emission spectrum of the system 
for a given input of the environment, such as a coherent drive.
We note that in the absence of the environment, the right-hand side of Eq.~\eqref{Eq.Heis_LangFour} is zero and the power spectrum vanishes, as it should.

\section{Memoryless approximation and Non-Hermitian effects} \label{Sec:2}

\begin{figure}[tbp] 
    \includegraphics[width=1\linewidth]{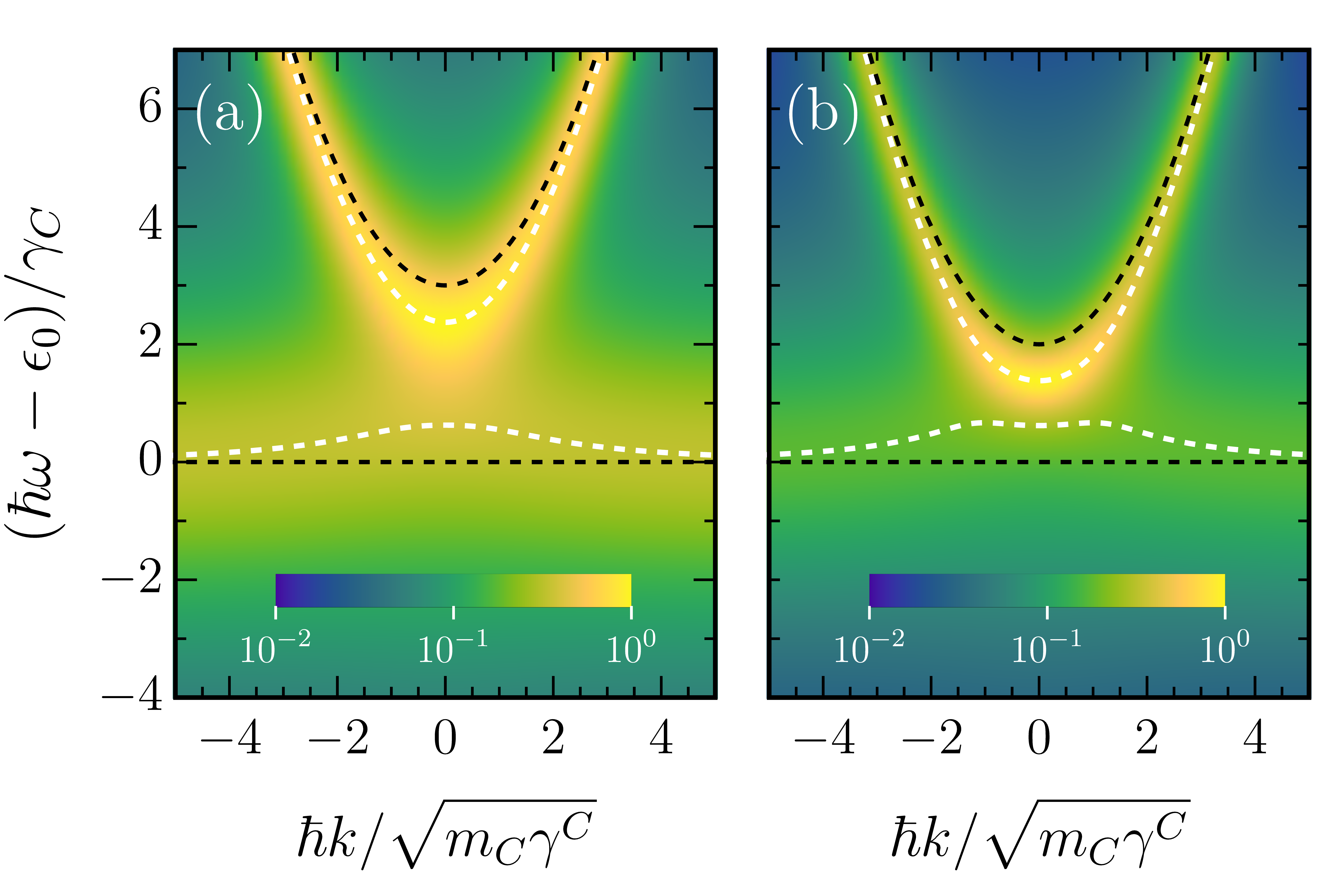}
\caption{Anomalous dispersions. The color plots show the power spectra $I(\k,\omega)$ (color scale in arbitrary units). The dashed white lines correspond to the real parts of Eq.~\eqref{Eq.eig} and the dashed-black lines represent the bare exciton and cavity photon kinetic energies $\epsilon_\k^X$ and $\epsilon_\k^C$. (a) $\delta/\gamma^C=3$, (b) $\delta/\gamma^C=2$.  The other parameters used for both panels are: $n_\k(\omega)=1$, $\gamma^X/\gamma^C=1.8$, $g_R/\gamma^X=0$, $m_C/m_X=0$.}
\label{fig:disp}
\end{figure}
The set of evolution equations in~\eqref{Eq.Heis_Lang_main} implies that the system operators $\hat{c}_\k,\hat{x}_\k$ at time $t$ are affected by their past at times $t'<t$ through the functions $\Gamma_\k^{AB}(\tau)$.
However, such memory effects disappear if we take the couplings $\kappa$ in Eq.~\eqref{eq:HamSE} to be independent of $q$, and approximate $\Gamma_\k^{CC}(\tau)\simeq \gamma_\k^C \delta(\tau) $, $\Gamma_\k^{XX}(\tau)\simeq \gamma_\k^X \delta(\tau) $, and $\Gamma_\k^{CX}(\tau)\simeq\sqrt{\gamma_\k^C \gamma_\k^X }\delta(\tau)$, where $\gamma_\k^C$ and $\gamma_\k^X$ correspond to the photon and exciton linewidths, respectively. \moved{These can be related to the model parameters via (see Appendix \ref{sec.AppBB})
\begin{align}\label{eq.damp2}
\gamma_\k^{C}&=\pi \rho_\k(\omega_0^X)|\kappa_\k^C|^2/\hbar,~~
\gamma_\k^{X}=\pi \rho_\k(\omega_0^X)|\kappa_\k^X|^2/\hbar.
\end{align}
where we have used $\rho_{\k}(\omega)\simeq\rho_{\k}(\omega_0^X)$, with $\omega_0^X=\epsilon_0/\hbar$} \ob{, and we have assumed that $\arg(\kappa_{\k}^{C*}\kappa_{\k}^{X})=0$, since the two parts of the system (cavity-photon and exciton) are located at the same place.}
Such a memoryless approximation is also referred to as the Markov approximation \cite{Gardiner1985}, and 
we expect it to be 
accurate when $\epsilon_\k^C,\epsilon_\k^X\gg \gamma_\k^C,\gamma_\k^X$.

Within this approximation,
 the matrix \eqref{Eq.Mmat} simplifies to
\begin{align} \label{Eq.Mmat2}
\mathcal{M}(\k,\omega)=\begin{pmatrix}
\hbar\omega-z_\k^C&&-\tilde{g}_\k \\
-\tilde{g}_\k&& \hbar\omega-z_\k^X
\end{pmatrix},
\end{align}
where we have introduced $z_\k^{C,X}=\epsilon_\k^{C,X}-i\gamma_\k^{C,X}$, and $\tilde{g}_\k=g_R-i\sqrt{\gamma_\k^{C}\gamma_\k^{X}}$. We can see that the matrix takes the form $\mathcal{M}(\k,\omega)=\hbar \omega \mathbb{1}- H_{\k}$, where $H_{\k}$ can be interpreted as an effective non-Hermitian Hamiltonian \cite{Ashida_nonHermitian}. Crucially, the imaginary part of the off-diagonal coupling is a consequence of the common environment, and would be absent in the typically assumed case of independent environments~\cite{Ciuti2006}.

Solving $\det[\mathcal{M}]=0$, one obtains the complex eigenvalues
\begin{align} \label{Eq.eig} 
\hbar\omega_\k^{L,U}&=\frac{1}{2}\left(z_\k^C+z_\k^X\pm \sqrt{(z_\k^C-z_\k^X)^2+4\tilde{g}_\k^2} \right).
\end{align} 
The power spectrum \eqref{Eq.Powerspect} can be calculated analytically and reads
\begin{align} \label{eq.PSmemoryless}
I(\k,\omega)&= \frac{A_\k(\omega)\gamma_\k^C+B_\k(\omega)\gamma_\k^X +D_\k(\omega)\sqrt{\gamma_\k^X\gamma_\k^C}}{|\hbar\omega-\hbar\omega_\k^L|^2|\hbar\omega-\hbar\omega_\k^U|^2} n_\k(\omega),
\end{align}
where $n_\k(\omega)=\pi \hbar\rho_{\k}(\omega_0^X) \langle \hat{e}_{\k,q_p}^{\text{I}\dagger}\hat{e}_{\k,q_p}^{\text{I}}\rangle$ is proportional to the photon distribution in the input field and we have defined
\begin{align} \nonumber
A_\k(\omega)&=4\left(|\tilde{g}_\k|^2+|\hbar\omega-z_\k^X|^2\right),\\ \nonumber
B_\k(\omega)&=4\left(|\tilde{g}_\k|^2+|\hbar\omega-z_\k^C|^2\right),\\ \nonumber D_\k(\omega)&=8\Re\left[(2\hbar\omega-z_\k^C-z_\k^X)\tilde{g}_\k^*\right].
\end{align}

We emphasize that 
$\gamma_\k^{C}$ and $\gamma_\k^{X}$ only account for 
radiative decay into the common photonic environment. In realistic systems, there can also exist nonradiative photonic and excitonic losses. As we show in Section \ref{Sec:3}, these allow for absorption to take place and can be incorporated into Eqs.~(\ref{Eq.Mmat2}--\ref{eq.PSmemoryless}) by adding additional decay rates in the definition of $z_\k^{C}, z_\k^{X}$. Importantly, the absorption spectrum can exhibit qualitatively similar features to the power spectrum. Here, we focus on the regime where such nonradiative losses are negligible with respect to $\gamma_\k^{X}$ and $\gamma_\k^{C}$. 
To shorten the notation in the discussion below, we introduce the $k$-dependent photon-exciton energy and linewidth detunings
$\delta_{\epsilon_\k}=\Re[z_\k^C-z_\k^X]$, and $\delta_{\gamma_\k}=-\Im[z_\k^C-z_\k^X]$.

\subsection{Level attraction and anomalous dispersions} 

In the limit of negligible Rabi coupling, the argument of the square root in Eq.~\eqref{Eq.eig} can give rise to level \textit{attraction} when $\delta_{\epsilon_\k}\neq0$ and $\gamma_\k^C\sim\gamma_\k^X$.
In particular, this effect provides a plausible and relatively simple explanation for recent puzzling photoluminescence measurements in semiconductor cavities \cite{Tawara2010,Dalacu2010,Valente2013,Dhara2018,Wurdack2023} which reported anomalous level attraction \footnote{Similar level attraction have also attracted  attention in cavity magnonic systems \cite{Harder2018,Wang2019}.}.  In the opposite limit of strong Rabi coupling, $g_R\gg \gamma_\k^C,\gamma_\k^X,\delta_{\epsilon_\k}$, the argument of the square root in Eq.~\eqref{Eq.eig} gives rise to level repulsion and to the conventional lower and upper polariton modes \cite{Weisbuch1992}.
Finally, we note that when $\gamma_\k^X/\gamma_\k^C\rightarrow0$, one recovers the case where the only radiative decay channel is from the cavity mode, as considered in Refs.~\cite{Sachdev1984,Barnett1986,Agarwal1986}.

In order to illustrate the phenomenon of level attraction and to highlight how this can, in turn, lead to anomalous dispersion relations in planar cavities, we have plotted in Fig.~\ref{fig:disp} two examples of power spectra. For the purposes of this illustration, 
we have used decay parameters independent of the in-plane wavevector $\gamma^{C,X}$ and a vanishing Rabi coupling $g_R/\gamma^{C,X}\simeq0$. %
The dispersion shown in panel (a) resembles the observation of Ref.~\cite{Dhara2018}, which reported an anomalous inverted parabolic behavior, i.e., a negative effective mass, of the lower line around $k=0$.
While this effect was originally believed to have a many-body origin \cite{Dhara2018}, it is not predicted by a recent more microscopic many-body theory \cite{Rana2021}. 
The dispersion displayed in panel (b) resembles that reported in Ref.~\cite{Wurdack2023}, in which the inverted parabolic behavior of the lower line was observed at $k\neq0$.
We note that the power spectrum calculated here is not strictly equivalent to an experimental photoluminescence spectrum, as in such measurements some relaxation and partial thermalization take place which 
tends to favor the occupation of the lower energy states. 

It is worth noticing that the single-mode version of the model in the regime $g_R/\gamma^{C,X}\simeq0$ \ob{is potentially relevant for semiconductor quantum-dot cavities. As such, a dissipative coupling mediated by the common photon environment could also have played a role for the level attraction reported in photoluminescence measurements in these systems \cite{Tawara2010,Dalacu2010,Valente2013}. In particular, we note that the observations of Ref.~\cite{Tawara2010} could not be fully explained by their model, and that the paper evokes some `unknown mode-pulling effects'.}
We emphasize that the present mechanism is solely due to the common photonic environment, which was not considered in \cite{Tawara2010,Dalacu2010,Valente2013,Dhara2018,Wurdack2023}, and it does not rely on material or temperature-dependent properties of solid-state emitters, nor on the pumping procedure. 

\subsection{Exceptional points and bound states in the continuum} 

\begin{figure}[tbp] 
    \includegraphics[width=1\linewidth]{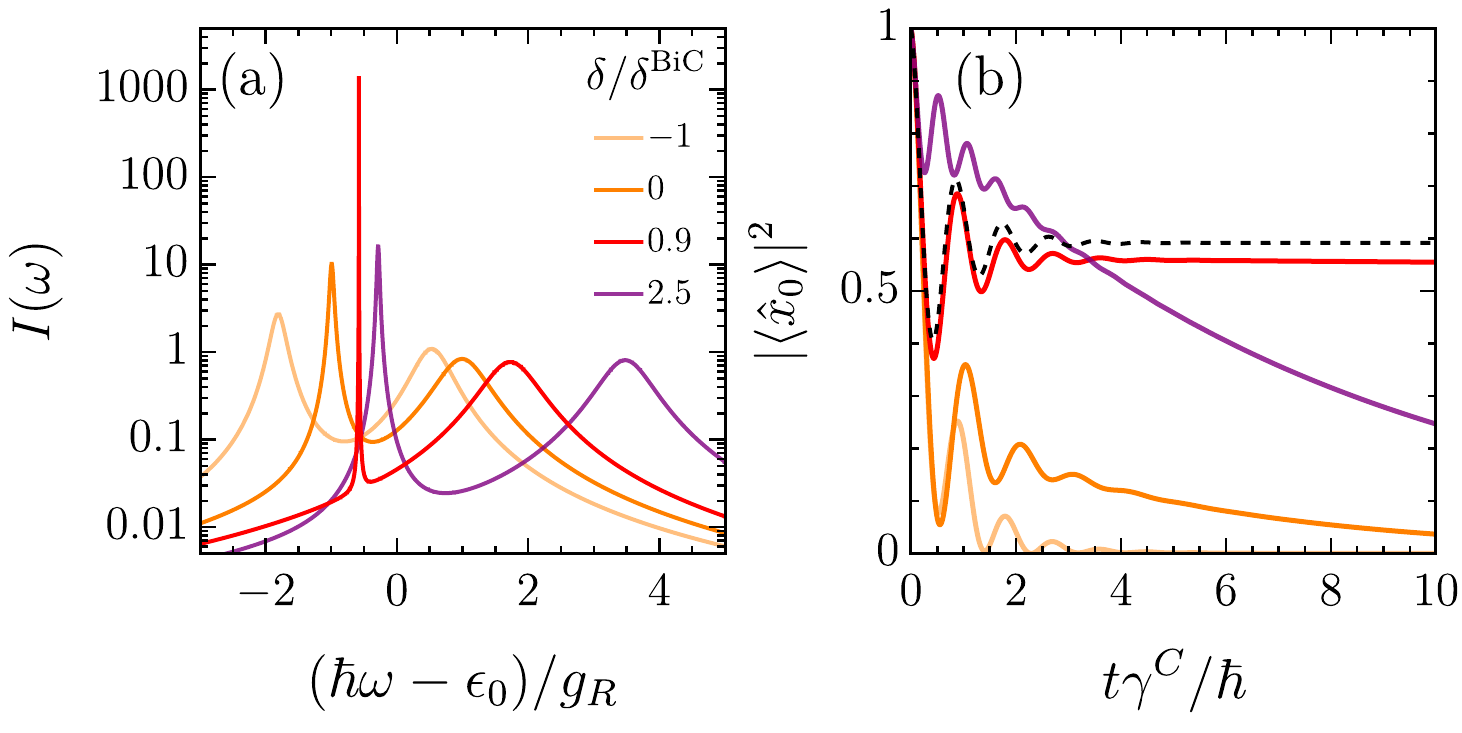}
\caption{Vicinity of a bound state in the continuum. (a) Power spectra for different detunings. (b) Dynamics of $|\langle\hat{x}_0\rangle|^2$ with initial conditions $\langle\hat{x}_0(0)\rangle=1$, $\langle\hat{c}_0(0)\rangle=0$ and vacuum environment. The solid colored lines correspond to the same detunings used in panel (a) and the dashed black line corresponds to $\delta=\delta^{\text{BiC}}$ given in Eq. \eqref{Eq.bictdepX}. The other parameters used in both panels are $g_R/\gamma^C=3$, $\gamma^X/\gamma^C=0.3$, $k=0$.}
\label{fig:BIC}
\end{figure}

Aside from the connection with recent experiments, the present model embeds additional interesting special cases. 
First, we note that exceptional points \cite{Berry2004,Heiss_2012,Ali2019} can arise when the square root in Eq. \eqref{Eq.eig} vanishes \footnote{The present exceptional points differ from those recently investigated in Refs.~\cite{Richter2019,Su2021,Krol2022} which are related to the polarization degree of freedom of cavity photons in anisotropic planar cavities. They also differ from the ones investigated theoretically in \cite{Hanai2019,Khurgin:2020} which require the presence of electronic pumping.}. 
This occurs when the following conditions for the photon-exciton detunings are both satisfied:
\begin{align}\label{eq.excep} 
\delta_{\epsilon_\k}^{\text{EP}}&=\pm2\sqrt{\gamma_\k^C\gamma_\k^X}, ~~\delta_{\gamma_\k}^{\text{EP}}=\mp2 g_R.
\end{align}
We note that these conditions would remain valid in the presence of an additional exciton broadening once it is incorporated into $\delta_{\gamma_\k}$; and that when $\gamma_\k^X=0$, Eq.~\eqref{eq.excep} corresponds to the condition for the so-called weak to strong coupling crossover \cite{Savona1995,Andreani1999}. EPs at such a crossover have been reported recently in Ref.~\cite{Gao2018} which relied on a polarization dependent Rabi splitting.
In our case, the Rabi coupling is constant and we have an in-plane isotropy such that Eq. \eqref{eq.excep} can give rise to rings of EPs which, to leading order in $m_C/m_X\ll1$, occur at $\hbar k_{\text{EP}}=\sqrt{2m_C(\delta_{\epsilon_\k}^{\text{EP}}-\delta)}$ \footnote{Rings of EPs are sometimes called ``exceptional rings'' \cite{BergholtzRMP2021}, an example of such a ring has been reported experimentally 
using photonic crystal slabs \cite{Zhen2015}.}.

Another interesting configuration appears when $\delta_{\epsilon_\k}=\delta_{\epsilon_\k}^{\text{BiC}}\equiv g_R\delta_{\gamma_\k}/\sqrt{\gamma_\k^C\gamma_\k^X}$, in which case the imaginary part of $\omega_\k^L$ in Eq.~\eqref{Eq.eig} vanishes exactly. As a consequence, the corresponding modes remain undamped \footnote{
We note that the hybrid light-matter BiCs presented here differ from the theoretical proposals based on a single emitter interacting with a semi-infinite 1D photonic environment \cite{Longhi2007,Tufarelli2013,Ciccarello2019,Barkemeyer2021}. Indeed, the existence of BiCs in our model does not rely on its semi-infinite nature, but on the presence of two resonances. 
We also note that the present BiCs are different from that investigated in the recent experiment of Ref.~\cite{Ardizzone2022}. In that work, the BiC originates from a periodically patterned waveguide structure and can exist in the absence of excitonic layer.}, 
which corresponds to the realization of bound states in the continuum \cite{Friedrich1985,hsu_bound_2016} (i.e., these states remain localized within the cavity despite the fact that they coexist with the continuum of photonic modes outside the cavity). These arise from the interference between two resonances (exciton and cavity photon) and are thus hybrid light-matter analogs of the BiCs proposed in Refs.~\cite{Fonda1960,Friedrich1985}.
Like the EPs above, the condition $\delta_{\epsilon_\k}=\delta_{\epsilon_\k}^{\text{BiC}}$ in a planar cavity can give rise to a ring of BiCs occuring at $\hbar k_{\text{BiC}}=\sqrt{2m_C(\delta_{\epsilon_\k}^{\text{BiC}}-\delta)}$. To our knowledge, the possibility to achieve such a ring is a novelty of the present isotropic system with respect to previous BiC realisations in other platforms \cite{hsu_bound_2016,Azzam2021review}.

In Fig.~\ref{fig:BIC}(a), we have plotted the power spectrum at $k=0$ for different values of $\delta/\delta^{\text{BiC}}$. We can see that as the detuning approaches $\delta_{}^{\text{BiC}}$ the width of the lower energy peak decreases and its amplitude increases. At exactly $\delta_{}=\delta_{}^{\text{BiC}}$, the width vanishes and the power spectrum \eqref{eq.PSmemoryless} diverges 
as $(\omega-\omega _0^{L})^{-2}$ \footnote{We note that an additional excitonic broadening would remove the exact cancellation of the width.}.
    To illustrate the impact of the BiC condition on the dynamics, in Fig.~\ref{fig:BIC}(b), we have plotted the amplitude of the exciton field as a function of time for the corresponding detunings. In this figure, we have used the solution of the memoryless equations of motion given in Appendix \ref{sec.AppB} with $\langle\hat{x}_0(t=0)\rangle=1$, and the environment and cavity mode in their vacuum state as initial conditions. We can observe damped oscillations at short times, while at long times, we see that the closer $\delta$ is to $\delta_{}^{\text{BiC}}$, the lower is the decaying slope. At $\delta_{}=\delta_{}^{\text{BiC}}$, the time-dependent amplitudes can be expressed as
\begin{subequations}\label{Eq.bictdep} 
\begin{align} 
|\langle \hat{c}_0(t)\rangle^{}|^2&=\frac{1+e^{-\frac{2 \gamma}{\hbar}t}-2e^{-\frac{ \gamma}{\hbar}t}\cos[\frac{\tilde{\Omega}}{\hbar}t]}{\gamma^2}\gamma^C\gamma^X,\\ \label{Eq.bictdepX} 
|\langle \hat{x}_0(t)\rangle^{}|^2&=\frac{\frac{\gamma^C}{\gamma^X}+\frac{\gamma^X}{\gamma^C}e^{-\frac{ 2\gamma}{\hbar}t}+2e^{-\frac{ \gamma}{\hbar}t}\cos[\frac{\tilde{\Omega}}{\hbar}t]}{\gamma^2}\gamma^C\gamma^X,
\end{align}
\end{subequations}
with $\gamma=\gamma^C+\gamma^X$ and $\tilde{\Omega}=g_R\gamma/\sqrt{\gamma^C\gamma^X}$.
We can clearly see that the amplitudes remain finite as $t\rightarrow \infty$ when $\gamma^C,\gamma^X\neq0$. In other words, the probabilities to find the emitter or the cavity excited do not vanish in the long time limit, which contrasts with the conventional behavior of damped vacuum Rabi oscillations in the absence of a common photonic environment.

\section{Additional losses, Reflection and Absorption} \label{Sec:3}

In the presence of a single photonic environment, one cannot formally describe absorption or reflection. Indeed, in the absence of any other loss channels, all input photons eventually return into the same environment, and thus the reflection is necessarily unity.
Therefore, to describe the absorption and reflection within the input-output formalism, one must include some additional losses in the model.

In realistic planar semiconductor cavities, there are several sources of nonradiative losses. For example, inhomogeneities of the mirrors can cause scattering of cavity photons into guided modes, while lattice phonons can cause scattering of excitons into nonradiative exciton states.
We will model these by adding two \textit{independent} baths with which the exciton and the cavity photon can interact
\begin{subequations}\label{eq:HamAddBath}
\begin{align} 
 \hat{H}_{m}&= \int dq\sum_{\mathbf{k}} \left[\epsilon_{\k,q}^{m}  \hat{m}_{\k q}^\dagger \hat{m}_{\k q} +\left( \kappa_{\k,q}^{m}  \hat{m}_{\k q}^\dagger \hat{x}_{\k}+ h.c.\right)\right] ,\\ 
  \hat{H}_{\gamma}&= \int dq\sum_{\mathbf{k}} \left[\epsilon_{\k,q}^{\gamma}  \hat{\gamma}_{\k q}^\dagger \hat{\gamma}_{\k q}+\left( \kappa_{\k,q}^{\gamma}  \hat{\gamma}_{\k q}^\dagger \hat{c}_{\k}+ h.c.\right)\right].
\end{align}
\end{subequations}
Here, $\epsilon_{\k,q}^{m}$ and $\epsilon_{\k,q}^{\gamma}$ correspond to the single particle energies in the distinct matter and photon baths respectively, and $\hat{m}_{\k q}$ and $\hat{\gamma}_{\k q}$ denote their annihilation operators.

The derivation provided in the Section \ref{Sec:1} can be straightforwardly generalized to include these additional baths, and one obtains the modified matrix equation
\begin{equation} \label{Eq.Heis_LangFour2}
\mathcal{M}'(\k,\omega)
\begin{pmatrix} 
\hat{\mathcal{C}}_\k(\omega)\\
\hat{\mathcal{X}}_\k(\omega)
 \end{pmatrix} = \begin{pmatrix} 
\hat{\mathcal{F}}_{\k}^C(\omega)+\hat{\mathcal{F}}_{\k}^{\gamma}(\omega)\\
\hat{\mathcal{F}}_{\k}^X(\omega)+\hat{\mathcal{F}}_{\k}^{m}(\omega)
 \end{pmatrix} ,
\end{equation}
with
\begin{align} \label{Eq.Mmatsup2}
\mathcal{M}'(\k,\omega)=\mathcal{M}(\k,\omega)+i \begin{pmatrix}
\tilde{\Gamma}_\k^{\gamma}(\omega)&&0\\
0&& \tilde{\Gamma}_\k^{m}(\omega)
\end{pmatrix}.
\end{align}
We can see that Eq.~\eqref{Eq.Heis_LangFour2} has a similar structure as Eq.~\eqref{Eq.Heis_LangFour}, with the presence of the additional baths included in $\tilde{\Gamma}_\k^{m}(\omega)$, $\tilde{\Gamma}_\k^{\gamma}(\omega)$  and $\hat{\mathcal{F}}_{\k}^{m}(\omega)$, $\hat{\mathcal{F}}_{\k}^{\gamma}(\omega)$.
In particular, we note that it gives rise to additional complex terms on the diagonal of the matrix $\mathcal{M}'$ with respect to $\mathcal{M}$.
Similarly to Eq. \eqref{Eq.Gamma_fourier}, these additional terms read
\begin{subequations}\label{Eq.Gamma_fourieradd}
\begin{align} 
\tilde{\Gamma}_\k^{m}(\omega)&=\frac{\pi}{\hbar}\int dq \,  |\kappa_{\k,q}^{m}|^2 \delta(\omega-\omega_{\k,q}^{m})
+\frac{i}{\hbar}\mathcal{P}\int dq \frac{|\kappa_{\k,q}^{m}|^2 }{\omega-\omega_{\k,q}^{m}},\\
\tilde{\Gamma}_\k^{\gamma}(\omega)&=\frac{\pi}{\hbar}\int dq \,  |\kappa_{\k,q}^{\gamma}|^2 \delta(\omega-\omega_{\k,q}^{\gamma})
+\frac{i}{\hbar}\mathcal{P}\int dq \frac{|\kappa_{\k,q}^{\gamma}|^2 }{\omega-\omega_{\k,q}^{\gamma}},
\end{align}
\end{subequations}
with $\omega_{\k,q}^{m}=\epsilon_{\k,q}^{m}/\hbar$, and $\omega_{\k,q}^{\gamma}=\epsilon_{\k,q}^{\gamma}/\hbar$.
In the same way as Eq.~\eqref{Eq.Force_fourier}, the operators $\hat{\mathcal{F}}_{\k}^{m}(\omega)$, $\hat{\mathcal{F}}_{\k}^{\gamma}(\omega)$ are given by
\begin{subequations}
\begin{align} 
\hat{\mathcal{F}}_{\k}^{m}(\omega)&=2\pi\kappa_{\k,q_m}^{m*}\rho_{\k}^{m}(\omega)\hat{m}_{\k,q_m}^{\text{I}} 
, \\
\hat{\mathcal{F}}_{\k}^{\gamma}(\omega)&=2\pi\kappa_{\k,q_\gamma}^{\gamma*}\rho_{\k}^{\gamma}(\omega)\hat{\gamma}_{\k,q_\gamma}^{\text{I}} . 
\end{align}
\end{subequations}
Here, however, $\rho_{\k}^{m}(\omega)$, $\rho_{\k}^{\gamma}(\omega)$ correspond to the density of states of the additional baths, and $\hat{m}_{\k q}^{\text{I}}$, $\hat{\gamma}_{\k q}^{\text{I}}$ denote their respective input operators, and $q_m$, $q_\gamma$ are defined by the resonance conditions $\omega = \omega^m_{\k,q_m}$ and $\omega = \omega^\gamma_{\k,q_\gamma}$.
\subsection{Generalized input-output relations} 
Due to the presence of the additional baths, the input-output relations are also modified. These can be expressed in a matrix form as
\begin{align} \label{Eq.InoutFin2m}
\begin{pmatrix}
\hat{e}_{\k,q_p}^{\text{O}}\\
\hat{m}_{\k,q_m}^{\text{O}}\\
\hat{\gamma}_{\k,q_\gamma}^{\text{O}}
\end{pmatrix}=
\mathcal{S}(\k,\omega)
\begin{pmatrix}
\hat{e}_{\k,q_p}^{\text{I}}\\
\hat{m}_{\k,q_m}^{\text{I}}\\
\hat{\gamma}_{\k,q_\gamma}^{\text{I}}
\end{pmatrix}.
\end{align}
 $\mathcal{S}(\k,\omega)$ is a $3\times3$ matrix, and its elements are
\begin{widetext}
\begin{subequations}\label{Eq.S2}
\begin{align} 
\mathcal{S}_{11}(\k,\omega)&=1-i\frac{2\pi\rho_{\k}(\omega)}{\hbar}\bigg(\left[\mathcal{M}'(\k,\omega)\right]_{11}^{-1} |\kappa_{\k,q_p}^{C}|^2+\left[\mathcal{M}'(\k,\omega)\right]_{22}^{-1} |\kappa_{\k,q_p}^{X}|^2 \nonumber \\ & ~~~~~~~~~~~~~~~~~~~~~~~~  +\left[\mathcal{M}'(\k,\omega)\right]_{12}^{-1} \kappa_{\k,q_p}^{C}\kappa_{\k,q_p}^{X*}+\left[\mathcal{M}'(\k,\omega)\right]_{21}^{-1} \kappa_{\k,q_p}^{X}\kappa_{\k,q_p}^{C*} \bigg),\\
\mathcal{S}_{22}(\k,\omega)&=1-i\frac{2\pi\rho_{\k}^{m}(\omega)}{\hbar}\left[\mathcal{M}'(\k,\omega)\right]_{22}^{-1} |\kappa_{\k,q_m}^{m}|^2,\\
\mathcal{S}_{33}(\k,\omega)&=1-i\frac{2\pi\rho_{\k}^{\gamma}(\omega)}{\hbar}\left[\mathcal{M}'(\k,\omega)\right]_{11}^{-1} |\kappa_{\k,q_\gamma}^{\gamma}|^2,\\
\mathcal{S}_{12}(\k,\omega)&=-i\frac{2\pi\rho_{\k}^{m}(\omega)}{\hbar}\left(\left[\mathcal{M}'(\k,\omega)\right]_{12}^{-1} \kappa_{\k,q_p}^{C}\kappa_{\k,q_m}^{m*}+\left[\mathcal{M}'(\k,\omega)\right]_{22}^{-1} \kappa_{\k,q_p}^{X}\kappa_{\k,q_m}^{m*}\right),
\\
\mathcal{S}_{13}(\k,\omega)&=-i\frac{2\pi\rho_{\k}^{\gamma}(\omega)}{\hbar}\left(\left[\mathcal{M}'(\k,\omega)\right]_{11}^{-1} \kappa_{\k,q_p}^{C}\kappa_{\k,q_\gamma}^{\gamma*}+\left[\mathcal{M}'(\k,\omega)\right]_{21}^{-1} \kappa_{\k,q_p}^{X}\kappa_{\k,q_\gamma}^{\gamma*}\right),
\\
\mathcal{S}_{21}(\k,\omega)&=-i\frac{2\pi\rho_{\k}(\omega)}{\hbar}\left(\left[\mathcal{M}'(\k,\omega)\right]_{21}^{-1} \kappa_{\k,q_p}^{C*}\kappa_{\k,q_m}^{m}+\left[\mathcal{M}'(\k,\omega)\right]_{22}^{-1} \kappa_{\k,q_p}^{X*}\kappa_{\k,q_m}^{m}\right),
\\
\mathcal{S}_{31}(\k,\omega)&=-i\frac{2\pi\rho_{\k}(\omega)}{\hbar}\left(\left[\mathcal{M}'(\k,\omega)\right]_{11}^{-1} \kappa_{\k,q_p}^{C*}\kappa_{\k,q_\gamma}^{\gamma}+\left[\mathcal{M}'(\k,\omega)\right]_{12}^{-1} \kappa_{\k,q_p}^{X*}\kappa_{\k,q_\gamma}^{\gamma}\right)
\\
\mathcal{S}_{23}(\k,\omega)&=-i\frac{2\pi\rho_{\k}^{\gamma}(\omega)}{\hbar}\left[\mathcal{M}'(\k,\omega)\right]_{21}^{-1} \kappa_{\k,q_p}^{\gamma*}\kappa_{\k,q_m}^{m},
\\
\mathcal{S}_{32}(\k,\omega)&=-i\frac{2\pi\rho_{\k}^{m}(\omega)}{\hbar}\left[\mathcal{M}'(\k,\omega)\right]_{12}^{-1} \kappa_{\k,q_m}^{m*}\kappa_{\k,q_\gamma}^{\gamma}.
\end{align}
\end{subequations}
\end{widetext}
We note that in the absence of the common environment (i.e., when $\kappa_{\k,q}^{X}=0$ and $\kappa_{\k,q}^{C}=0$), one recovers the input-output relation obtained in Ref.~\cite{Ciuti2006} in the case of two independent baths for the cavity photons and the excitons. However, as we  mentioned above, the transformation $\mathcal{S}$ cannot be unitary when the different baths are not equivalent.

\subsection{Reflection and Absorption}

\begin{figure}[tbp] 
    \includegraphics[width=1\linewidth]{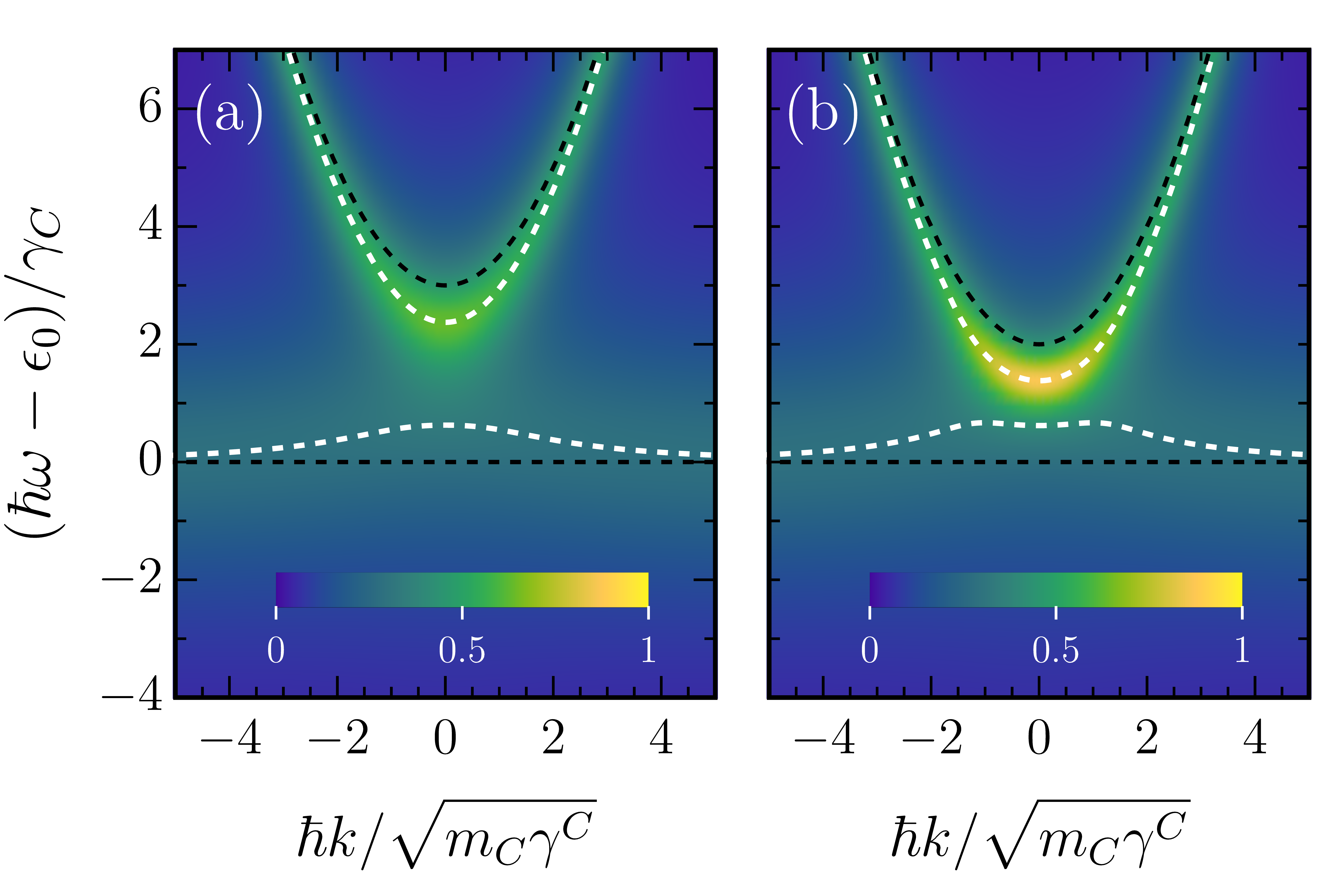}
\caption{Absorption spectra. The color plots are obtained from Eq.~\eqref{eq.Ab_markov}. 
The dashed white lines correspond to the real parts of the eigenvalues [Eq.~\eqref{Eq.eig}] and the dashed-black lines represent the bare exciton and cavity photon kinetic energies $\epsilon_\k^X$ and $\epsilon_\k^C$. In both panels, the nonradiative decay parameters are $\gamma^{\gamma}/\gamma^C=\gamma^{m}/\gamma^C=0.15$. All the others parameters are identical to the ones used in Figure \ref{fig:disp}.}
\label{fig:abs}
\end{figure}

In order to introduce the notions of reflection and absorption one needs to use the conservation of the total particle number in the baths. This is given by the relation

\begin{widetext}
\begin{align}\label{eq.Iconservedm}
&\int d\omega \left[\rho_\k(\omega)\langle\hat{e}_{\k,q_p}^{\text{O}\dagger}\hat{e}_{\k,q_p}^{\text{O}}\rangle+ \rho_\k^{m}(\omega)\langle\hat{m}_{\k,q_m}^{\text{O}\dagger}\hat{m}_{\k,q_m}^{\text{O}}\rangle+\rho_\k^{\gamma}(\omega)\langle\hat{\gamma}_{\k,q_\gamma}^{\text{O}\dagger}\hat{\gamma}_{\k,q_\gamma}^{\text{O}}\rangle\right] \nonumber \\
& ~~~= \int d\omega \left[\rho_\k(\omega) \langle\hat{e}_{\k,q_p}^{\text{I}\dagger}\hat{e}_{\k,q_p}^{\text{I}}\rangle+ \rho_\k^{m}(\omega) \langle\hat{m}_{\k,q_m}^{\text{I}\dagger}\hat{m}_{\k,q_m}^{\text{I}}\rangle+ \rho_\k^{\gamma}(\omega)\langle\hat{\gamma}_{\k,q_\gamma}^{\text{I}\dagger}\hat{\gamma}_{\k,q_\gamma}^{\text{I}}\rangle \right].
\end{align}
\end{widetext}
The fact that the three baths are different is accounted for by the density of states.
In a reflection/absorption experiment, one excites the system from the photonic environment outside of the cavity while the other baths are initially in their vacuum states; hence, one has  $ \langle\hat{e}_{\k,q_p}^{\text{I}\dagger}\hat{e}_{\k,q_p}^{\text{I}}\rangle\neq0$ and the other input averages are zero.
One can then use Eq.~\eqref{Eq.InoutFin2m} to express the output averages in term of the input ones such that Eq.~\eqref{eq.Iconservedm} reduces to
\begin{align}
\rho_\k(1-|\mathcal{S}_{11}(\k,\omega)|^2)-|\mathcal{S}_{21}(\k,\omega)|^2\rho_\k^{m}-|\mathcal{S}_{31}(\k,\omega)|^2\rho_\k^{\gamma}=0,\nonumber
\end{align}
from which we can deduce the reflection and absorption spectra
\begin{subequations}
\begin{align}\label{eq.RA}
\mathcal{R}(\k,\omega)&=|\mathcal{S}_{11}(\k,\omega)|^2, \\\mathcal{A}(\k,\omega)&= |\mathcal{S}_{21}(\k,\omega)|^2\frac{\rho_\k^{m}}{\rho_\k}+|\mathcal{S}_{31}(\k,\omega)|^2\frac{\rho_\k^{\gamma}}{\rho_\k}.
\end{align}
\end{subequations}
We note that the relation $\mathcal{R}+\mathcal{A}=1$ is respected, as it should, in the absence of any transmission.
(We recall that no transmission can occur in the scenario we consider since only one mirror is not perfect, as schematized in Figure \ref{fig1}(a)).

Within the memoryless approximation, the absorption spectrum can be expressed as
\begin{align}\label{eq.Ab_markov}
\mathcal{A}(\k,\omega)=\gamma_\k^{\gamma}A_\gamma(\k,\omega)+\gamma_\k^{m} A_m(\k,\omega),
\end{align}
with
\begin{subequations}\label{eq.Ab}
    \begin{align}
A_\gamma(\k,\omega)&=4 \frac{\gamma_\k^{C} |\hbar\omega-z_\k^X|^2 +\gamma_\k^{X} |\tilde{g}_\k|^2}{|\hbar\omega-z_\k^L|^2|\hbar\omega-z_\k^U|^2} \\
& ~~~ +8 \sqrt{\gamma_\k^{X} \gamma_\k^{C} }  \frac{ \Re\left[\tilde{g}_\k^* (\hbar\omega-z_\k^X)\right]}{|\hbar\omega-z_\k^L|^2|\hbar\omega-z_\k^U|^2}, \nonumber
\\
A_m(\k,\omega)&=4\frac{\gamma_\k^{C} |\tilde{g}_\k|^2+\gamma_\k^{X} |\hbar\omega-z_\k^C|^2 }{|\hbar\omega-z_\k^L|^2|\hbar\omega-z_\k^U|^2}\\
&~~~ +8\sqrt{\gamma_\k^{X} \gamma_\k^{C} } \frac{ \Re\left[\tilde{g}_\k^* (\hbar\omega-z_\k^C)\right]}{|\hbar\omega-z_\k^L|^2|\hbar\omega-z_\k^U|^2}. \nonumber
\end{align}
\end{subequations}
Here, we have $z_\k^X=\epsilon_\k^X-i \gamma_\k^X-i \gamma_\k^m$ and $z_\k^C=\epsilon_\k^C-i \gamma_\k^C-i \gamma_\k^\gamma$, and the additional photon and matter decay parameters $\gamma_\k^\gamma,\gamma_\k^m$ are defined in a similar way as $\gamma_\k^X$ and $\gamma_\k^C$ with 
\begin{align} 
\gamma_\k^{m}&=\pi \rho_\k^m(\omega_0^X)|\kappa_\k^m|^2/\hbar,~~
\gamma_\k^{\gamma}=\pi \rho_\k^\gamma(\omega_0^X)|\kappa_\k^\gamma|^2/\hbar.
\end{align}

Since in an absorption experiment the input is from the common photonic environment, the analytical expression we obtained previously for the power spectrum  \eqref{eq.PSmemoryless} remains valid
once the additional decays are included in $z_\k^X$ and $z_\k^C$. 
One can then observe that the power spectrum is related to the absorption spectrum as
    \begin{align}
I(\k,\omega)=\left(A_\gamma(\k,\omega)+A_m(\k,\omega)\right)n_\k(\omega).
\end{align}
Interestingly, we can see that the contributions from $A_\gamma(\k,\omega)$ and $A_m(\k,\omega)$ are weighted by the nonradiative loss strengths in the absorption spectrum \eqref{eq.Ab_markov} but not in the power spectrum. This difference is related to the fact that there are two independent channels for absorption to take place, while the power spectrum can be defined in the absence of these nonradiative channels.

In Figure~\ref{fig:abs}, we have plotted the absorption spectra using the parameters used in Figure \ref{fig:disp} and symmetric nonradiative decay parameters $\gamma^{\gamma}/\gamma^C=\gamma^{m}/\gamma^C=0.15$. Apart from the different color scale, we can see that the absorption spectrum exhibits similar features as the power spectrum, as expected.
This demonstrates that, in principle, one can experimentally access the anomalous dispersions in absorption measurements.

\section{Conclusions} \label{Sec:Conc}
We have demonstrated that the presence of a common photonic environment can lead to an effective dissipative coupling between light and matter in a nonideal cavity. Using a memoryless approximation, we have obtained analytical results for the power spectrum and the complex eigenenergies of this open system. This allowed us to highlight a potential connection with recent experiments that have reported anomalous level attraction, and to provide simple conditions under which rings of EPs and BiCs are expected.
\ob{Furthermore, we have extended the model to incorporate non-radiative losses, thereby demonstrating that level attraction and anomalous dispersions could be probed in absorption experiments.}

Our results open up intriguing perspectives. The inclusion of nonlinearities, which are system-dependent, could unveil novel regimes to generate photon antibunching in cavity systems \cite{Imamoglu1997,Verger2006,Liew2010,bamba2011origin,Gullans2013}. Furthermore, to our knowledge, the possibility for the emitters and the cavity modes to interact with a common photonic environment is not accounted for in the usual models for single photon sources \cite{Auffeves2009,Grange2015,Choi2019} or laser theories \cite{haken1985laser}
, and it could be interesting to investigate whether it affects some of their properties. In this context, one expects that the hybrid light-matter BiC condition could favor low threshold lasing as evidenced with photonic BiCs \cite{hsu_bound_2016,Azzam2021review,Kodigala2017}.

\begin{acknowledgments}
We gratefully acknowledge D\oldk abr\' owka Biega\' nska and Maciej Pieczarka for fruitful exchanges that brought the problem of anomalous dispersions to our attention. We also thank Matthias Wurdack, Elena Ostrovskaya, and Eliezer Estrecho for useful discussions.
We acknowledge support from the Australian Research Council Centre of Excellence in Future Low-Energy Electronics Technologies (CE170100039).
JL and MMP are also supported through Australian Research Council Future Fellowships FT160100244 and FT200100619, respectively. KC acknowledges support from an Australian Government Research Training Program (RTP) Scholarship.
\end{acknowledgments}

\appendix

\begin{widetext}

\section{Connection to the Green's function formalism}\label{sec.AppBBA}
In this Appendix, we highlight the connection between the matrix $\mathcal{M}$ derived in the main text and the Green's matrix of the system in presence of a common environment.
The system single particle retarded Green's matrix can be defined as
\begin{align}\label{Eq.Green_syst}
\bold{G}(\k,t)=\begin{pmatrix}
G_{CC}(\k,t)&&G_{CX}(\k,t)\\
G_{XC}(\k,t)&& G_{XX}(\k,t)
\end{pmatrix}=-\frac{i}{\hbar}\theta(t)\begin{pmatrix}
\langle\hat{c}_\k(t)\hat{c}_\k^{\dagger}(0) \rangle&&\langle\hat{c}_\k(t)\hat{x}_\k^{\dagger}(0) \rangle\\
\langle\hat{x}_\k(t)\hat{c}_\k^{\dagger}(0) \rangle && \langle\hat{x}_\k(t)\hat{x}_\k^{\dagger}(0) \rangle
\end{pmatrix},
\end{align}
where the averages are taken over the vacuum.
The cavity photon and exciton Green's functions correspond to the diagonal elements of this matrix.
It is also convenient to introduce the following system-environment retarded Green's functions 
\begin{subequations} \label{Eq.Green_systenv}
\begin{align}
G_{e C}(\k,q,t)&=-\frac{i}{\hbar}\theta(t)\langle\hat{e}_{\k q}(t)\hat{c}_\k^{\dagger}(0) \rangle,\\
G_{eX}(\k,q,t)&=-\frac{i}{\hbar}\theta(t)\langle\hat{e}_{\k q}(t)\hat{x}_\k^{\dagger}(0) \rangle.
\end{align}
\end{subequations}
Taking the time derivative of Eq.~\eqref{Eq.Green_syst} gives
\begin{align}\label{Eq.GreenpartialT}
i\hbar \partial_t \bold{G}(\k,t)&=\delta(t)\begin{pmatrix}
\langle\hat{c}_\k(t)\hat{c}_\k^{\dagger}(0) \rangle&&\langle\hat{c}_\k(t)\hat{x}_\k^{\dagger}(0) \rangle\\
\langle\hat{x}_\k(t)\hat{c}_\k^{\dagger}(0) \rangle && \langle\hat{x}_\k(t)\hat{x}_\k^{\dagger}(0) \rangle
\end{pmatrix} -\frac{i}{\hbar}\theta(t)\begin{pmatrix}
\langle  i\hbar \partial_t \hat{c}_\k(t)\hat{c}_\k^{\dagger}(0) \rangle&&\langle i\hbar \partial_t \hat{c}_\k(t)\hat{x}_\k^{\dagger}(0) \rangle\\
\langle i\hbar \partial_t \hat{x}_\k(t)\hat{c}_\k^{\dagger}(0) \rangle && \langle  i\hbar \partial_t \hat{x}_\k(t)\hat{x}_\k^{\dagger}(0) \rangle
\end{pmatrix}.
\end{align}
One can then use the equations of motion for the system operators to re-express the second term on the right hand side. 
Since the total Hamiltonian considered here only contains single-body couplings, the elements of this term can be expressed as linear combinations of single particle Green's functions.

Making use of the equations of motion for the system operators \eqref{Eq.Heis_sys}, and Fourier transformation, one can obtain
\begin{subequations} \label{Eq.Green_syst_fourier}
\begin{align}
\hbar\omega \mathcal{G}_{CC}(\k,\omega)&=\epsilon_\k^C \mathcal{G}_{CC}(\k,\omega) + g_R \mathcal{G}_{XC}(\k,\omega) +\int dq \, \kappa_{\k,q}^{C*}\mathcal{G}_{e C}(\k,q,\omega) +1,
\\
\hbar\omega \mathcal{G}_{XX}(\k,\omega)&=\epsilon_\k^X \mathcal{G}_{XX}(\k,\omega) + g_R \mathcal{G}_{CX}(\k,\omega) +\int dq \, \kappa_{\k,q}^{X*}\mathcal{G}_{e X}(\k,q,\omega)+1,
\\
\hbar\omega \mathcal{G}_{XC}(\k,\omega)&=\epsilon_\k^X \mathcal{G}_{XC}(\k,\omega) + g_R\mathcal{G}_{CC}(\k,\omega) +\int dq \, \kappa_{\k,q}^{X*}\mathcal{G}_{e C}(\k,q,\omega),
\\
\hbar\omega \mathcal{G}_{CX}(\k,\omega)&=\epsilon_\k^C \mathcal{G}_{CX}(\k,\omega) + g_R\mathcal{G}_{XX}(\k,\omega) +\int dq \, \kappa_{\k,q}^{C*}\mathcal{G}_{e X}(\k,q,\omega),
\end{align}
\end{subequations}
where $\mathcal{G}_{AB}$ denotes the Fourier transform of $G_{AB}$.
The same procedure applied to the system-environment Green's functions \eqref{Eq.Green_systenv} gives
\begin{subequations} \label{Eq.Green_systenv_fourier}
\begin{align}
\hbar \omega \mathcal{G}_{e C}(\k,q,\omega)&=\epsilon_{\k,q}^{E} \mathcal{G}_{e C}(\k,q,\omega) + \kappa_{\k, q}^C\mathcal{G}_{CC}(\k,\omega) + \kappa_{\k, q}^X\mathcal{G}_{XC}(\k,\omega), 
\\
\hbar \omega \mathcal{G}_{e X}(\k,q,\omega)&=\epsilon_{\k,q}^{E} \mathcal{G}_{eX}(\k,q,\omega) + \kappa_{\k, q}^C\mathcal{G}_{CX}(\k,\omega)+ \kappa_{\k, q}^X\mathcal{G}_{XX}(\k,\omega).
\end{align}
\end{subequations}
We can now inject the system-environment Green's function obtained from Eq. \eqref{Eq.Green_systenv_fourier} into Eq. \eqref{Eq.Green_syst_fourier} and obtain the matrix equation
\begin{align}\label{eq.Matrixequation}
\begin{pmatrix}
\hbar\omega- \epsilon_\k^C +i \tilde{\Gamma}_\k^{CC}(\omega)  &&-g_R+i \tilde{\Gamma}_\k^{CX}(\omega)  \\
-g_R+i \tilde{\Gamma}_\k^{XC}(\omega) && \hbar\omega- \epsilon_\k^X +i \tilde{\Gamma}_\k^{XX}(\omega)
\end{pmatrix}
\begin{pmatrix}
\mathcal{G}_{CC}(\k,\omega) &&\mathcal{G}_{CX}(\k,\omega)\\
\mathcal{G}_{XC}(\k,\omega)&& \mathcal{G}_{XX}(\k,\omega)
\end{pmatrix}=\mathbb{1},
\end{align}
where $\mathbb{1}$ is the identity matrix. 
In Eq. \eqref{eq.Matrixequation}, one can recognize the matrix $\mathcal{M}(\k,\omega)$ given in Eq.~\eqref{Eq.Mmat} of the main text, and observe that it corresponds to the inverse of the system retarded Green's matrix in frequency space.
\end{widetext}

\section{Memoryless approximation} \label{sec.AppBB}

The set of evolution equations in 
\eqref{Eq.Heis_Lang_main} shows that the system operators $\hat{c}_\k,\hat{x}_\k$ at time $t$ are affected by all their past at times $t'<t$. 
This influence is encoded in the functions $\Gamma_\k^{AB}(\tau)$ defined in Eq.~\eqref{Eq.gammaT}.
As mentioned in the main text, one can simplify these equations by using a memoryless approximation. 
One can motivate this approximation by 
inspecting the functions $\Gamma_\k^{AB}(\tau)$
\begin{eqnarray} 
\Gamma_\k^{AB}(\tau)&=&\theta(\tau)\frac{1}{\hbar}\int_{-\infty}^\infty dq\, \kappa_{\k,q}^{A*}\kappa_{\k,q}^{B}e^{-\frac{i}{\hbar}\epsilon_{\k,q}^{E}\tau}\\
&=&\theta(\tau)\frac{1}{\hbar}\int_0^\infty d\omega\, \rho_{\k}(\omega) \kappa_{\k,q_p}^{A*}\kappa_{\k,q_p}^{B}e^{-i\omega\tau}.
\end{eqnarray}
We expect the system-environment coupling strengths to be non-negligible only when the environment frequencies are in the vicinity of the system ones.
Since the relevant frequencies are large (these are in the vicinity of the emitter frequency $\omega_0^X=\epsilon_0/\hbar$),  
we can 
neglect the frequency dependence of $\kappa_{\k,q_p}^{A}\simeq \kappa_{\k}^{A}$ and take $\rho_{\k}(\omega)\simeq\rho_{\k}(\omega_0^X)$. Then one has 
\begin{align} 
&\int_{0}^\infty d\omega\, \rho_{\k}(\omega) \kappa_{\k,q_p}^{A*}\nonumber\kappa_{\k,q_p}^{B}e^{-i\omega\tau} \\ & ~~\simeq e^{-i\omega_0^X\tau}\int_{-\infty}^\infty d\omega\, \rho_{\k}(\omega_0^X) \kappa_{\k}^{A*}\kappa_{\k}^{B}e^{-i\omega\tau} \\
& ~~=2\pi\rho_{\k}(\omega_0^X)\kappa_{\k}^{A*}\kappa_{\k}^{B}\delta(\tau),
\end{align}
and thus one can approximate
\begin{equation}\label{eq.damp}
\Gamma_\k^{AB}(\tau)\simeq \frac{\pi}{\hbar}\rho_\k(\omega_0^X)\kappa_\k^{A*}\kappa_\k^B \delta(\tau) .
\end{equation}
 \ob{Assuming $\arg(\kappa_{\k}^{C*}\kappa_{\k}^{X})=0$,} in this approximation, Eqs.~\eqref{Eq.Heis_Lang_main} reduce to
\begin{subequations}\label{Eq.Heis_Lang_Meomoryless}
\begin{align} 
 i\hbar\partial_t\hat{c}_\k &=z_\k^{C}\hat{c}_\k + \tilde{g}_\k\hat{x}_\k+\hat{F}_\k^{C},\\ 
  i\hbar\partial_t\hat{x}_\k &=z_\k^{X} \hat{x}_\k  +\tilde{g}_\k \hat{c}_\k+\hat{F}_\k^{X},
\end{align}
\end{subequations}
with $z_\k^{C,X}=\epsilon_\k^{C,X}-i\gamma_\k^{C,X}$ and $\tilde{g}_\k=g_R-i\sqrt{\gamma_\k^{C}\gamma_\k^{X}}$ and the decay parameters
\begin{align}\label{eq.damp2sup}
\gamma_\k^{C}&=\pi \rho_\k(\omega_0^X)|\kappa_\k^C|^2/\hbar,~~
\gamma_\k^{X}=\pi \rho_\k(\omega_0^X)|\kappa_\k^X|^2/\hbar.
\end{align}
This leads to a simplification of the matrix \eqref{Eq.Mmat} 
into \eqref{Eq.Mmat2} 
in the main text.
In addition, within this approximation the force operators in frequency space \eqref{Eq.Force_fourier} read
\begin{subequations}\label{Eq.Force_fourier_memoryless}
\begin{align} 
\hat{\mathcal{F}}_{\k}^{C}(\omega)&=2\pi\kappa_{\k}^{C*}\rho_{\k}(\omega_0^X)\hat{e}_{\k,q_p}^{\text{I}} , 
\\ 
\hat{\mathcal{F}}_{\k}^{X}(\omega)&=2\pi\kappa_{\k}^{X*}\rho_{\k}(\omega_0^X)\hat{e}_{\k,q_p}^{\text{I}},
\end{align}
\end{subequations}
and one has
\begin{align} 
\langle\hat{\mathcal{F}}_{\k}^{A\dagger}(\omega)\hat{\mathcal{F}}_{\k}^{B}(\omega)\rangle=4\pi \hbar\rho_{\k}(\omega_0^X) \sqrt{\gamma_{\k}^{A}\gamma_{\k}^{B}} \langle \hat{e}_{\k,q_p}^{\text{I}\dagger}\hat{e}_{\k,q_p}^{\text{I}} \rangle.
\end{align}

\section{Field amplitude dynamics} \label{sec.AppB}
Assuming that the environment is initially in its vacuum and using the memoryless approximation, one can calculate analytically the evolution of the system field amplitudes. Taking the average of the evolution equations in \eqref{Eq.Heis_Lang_Meomoryless}, one has

\begin{subequations}\label{Eq.Heis_Lang_MeomorylessAV}
\begin{eqnarray} 
 i\hbar\partial_t\langle\hat{c}_\k \rangle&=&z_\k^C\langle\hat{c}_\k\rangle +\tilde{g}_\k\langle\hat{x}_\k\rangle,\\ 
  i\hbar\partial_t\langle\hat{x}_\k\rangle &=&z_\k^X \langle\hat{x}_\k\rangle  +\tilde{g}_\k \langle\hat{c}_\k\rangle.
\end{eqnarray}
\end{subequations}
The solutions are of the form
\begin{widetext}
\begin{subequations}\label{Eq.Heis_Lang_MeomorylessSol}
\begin{eqnarray} 
\langle\hat{c}_\k(t) \rangle&=\frac{1}{\hbar\omega_\k^U-\hbar\omega_\k^L} \{e^{-i\omega_\k^U t}\left[\langle\hat{x}_\k(0) \rangle\tilde{g}_\k+ \langle\hat{c}_\k(0) \rangle(\hbar\omega_\k^U-z_\k^X)\right]-e^{ -i\omega_\k^Lt}\left[\langle\hat{x}_\k(0) \rangle\tilde{g}_\k- \langle\hat{c}_\k(0) \rangle(\hbar\omega_\k^U-z_\k^C)\right]\},
\\
\langle\hat{x}_\k(t) \rangle&=\frac{1}{\hbar\omega_\k^U-\hbar\omega_\k^L} \{e^{-i\omega_\k^U t}\left[\langle\hat{c}_\k(0) \rangle\tilde{g}_\k+ \langle\hat{x}_\k(0) \rangle(\hbar\omega_\k^U-z_\k^C)\right]-e^{ -i\omega_\k^Lt}\left[\langle\hat{c}_\k(0) \rangle\tilde{g}_\k- \langle\hat{x}_\k(0) \rangle(\hbar\omega_\k^U-z_\k^X)\right]\}.
\end{eqnarray}
\end{subequations}
\end{widetext}
Using the initial conditions $\langle\hat{c}_\k(0) \rangle=0$ and $\langle\hat{x}_\k (0)\rangle=1$, one can obtain Eq.~\eqref{Eq.bictdep} 
 for $\delta=\delta^{\text{BiC}}$ and $k=0$.

It is interesting to note that the model used to fit the experimental Rabi oscillations reported in Ref.~\cite{Dominici2014} contains a phenomenological ``upper polariton decay/dephasing" term that seems to act as the present dissipative coupling ($\Im[\tilde {g}]$) in the evolution equations for the field amplitudes \eqref{Eq.Heis_Lang_MeomorylessAV}. This suggests that the mechanism introduced in our work could also have played a role in this time-resolved experiment.

\bibliography{biblio}
\end{document}